\documentclass[aps,prl,showpacs,preprintnumbers,nofootinbib,floatfix,twocolumn]{revtex4}

\usepackage{bm}
\usepackage{latexsym}
\usepackage{dcolumn}
\usepackage{amsfonts,amssymb}
\usepackage{graphicx,epsfig}
\usepackage{psfrag}

\begin{document}


\title{Casimir Effect confronts Cosmological Constant }
\author{Gaurang Mahajan}
\email{gaurang@iucaa.ernet.in}
\author{Sudipta Sarkar}
\email{sudipta@iucaa.ernet.in}
\author{T. Padmanabhan}
\email{paddy@iucaa.ernet.in}
\affiliation{IUCAA,
Post Bag 4, Ganeshkhind, Pune - 411 007, India\\}

\date{\today}


\begin{abstract}
It has been speculated that the zero-point energy of the vacuum, regularized due to the existence of a suitable ultraviolet cut-off scale, could be the source of the non-vanishing cosmological constant that is driving the present acceleration of the universe. We show that the presence of such a cut-off can significantly alter the results for the Casimir force between parallel conducting plates and even lead to repulsive Casimir force when the plate separation is smaller than the cut-off scale length. Using the current experimental data we rule out the possibility that the observed cosmological constant arises from the zero-point energy which is made finite by a suitable cut-off. Any such cut-off which is consistent with the observed Casimir effect will lead to an energy density which is about $10^{12}$ times larger than the observed one, if gravity couples to these modes. The implications are discussed.
\end{abstract}

\maketitle
\vskip 0.5 in
\noindent
The Casimir effect is one of the remarkable observable consequences of the existence of quantum fluctuations \cite{cas,milonni,milton,greiner}. In the standard approach of canonical field quantization, the total zero-point energy of the vacuum ($E_{0}$) is a divergent quantity. In the presence of macroscopic bodies (like a pair of conducting plates) leading to some non-trivial boundary conditions $(\partial \Gamma )$, the zero-point fluctuations of the vacuum get modified to some other value $E[\partial \Gamma ]$, with a finite difference
$
\Delta E = E[\partial \Gamma ] - E_{0}
$.
This change manifests as a tangible force between the macroscopic bodies which distort the vacuum. A well-studied example is the presence of an \textit{attractive} force between two neutral, perfectly conducting parallel plates \cite{cas, milonni, greiner}. In free space, the spectrum of possible wave modes of the vacuum forms a continuum; but the presence of plates restricts the normal components of the allowed wave modes between them to discrete values. This discreteness leads to a finite lowering of the vacuum energy, resulting in a force of attraction. This force has been measured to a high degree of precision in several experiments over the past decade \cite{expts}.

Any modification of the intrinsic spectrum of zero-point fluctuations will also leave an imprint on the Casimir effect. In particular, if there exists a cut-off length $L_{c}$ in nature, due to some unknown physics, it would act as a natural UV regulator, suppressing the field modes with wavelengths $\lambda \lesssim L_{c}$ and leading to a modification of the Casimir force between bodies. In the case of parallel plates separated by a distance $a$, for example, the maximum wavelength allowed between them is $\lambda_{max} \sim 2 a$, and it is the modes with wavelengths lying in the range $L_{c} \lesssim \lambda \lesssim 2 a$ which will contribute to the Casimir effect. This is indistinguishable from the standard result if $L_c\ll a$ --- which is usually the case since the cut-off lengths suggested in the literature \cite{minlenpap} are usually of the order of Planck length. But if this cut-off length is significantly larger,  there would be significant corrections to the standard result when $L_c\gtrsim a$ (see \cite{nouc} for related ideas). Presently, there is a strong experimental evidence for the electromagnetic Casimir force between metallic bodies up to separations around 100 nm and no significant deviations from the predicted results have been observed \cite{expts}. This suggests that any possible cut-off length scale $L_{c}$ must lie below this value. This translates into a lower bound on the vacuum energy density of $\rho_V\gtrsim 1$ $(eV)^{4}$.

There are some proposals in the literature \cite{beck} that the zero-point energy of vacuum --- made finite by a cut-off --- may contribute to gravity and lead to a non-vanishing cosmological constant responsible for the currently observed accelerated expansion of the universe. Current cosmological observations seem to favour cosmological constant as the candidate for dark energy \cite{cc} and constrain the  energy density contributed by it to be roughly $\rho_{DE} \approx 10^{-11} (eV)^{4}$, with a corresponding length scale of the order of 0.1 mm. The significant discrepancy of this value with the bound obtained from the Casimir effect measurements makes it unlikely that the cosmological constant has anything to do with zero-point energy rendered finite by a cut-off of unknown physical origin but having the correct length scale.

We put these ideas on a firmer footing by systematically studying the effect of a cut-off length scale ($L_{c}\sim k_{c}^{-1} $) on the Casimir energy evaluated in a $(d+1)$ dimensional spacetime. In the presence of such a cut-off, the contribution to the vacuum energy of any mode with frequency $k$ would get modified by an appropriate factor $f(k/k_{c})$, which, in general, would be a function with
$
 f(k/k_{c})\approx 1$ for $k \ll k_{c}$ and
$f(k/k_{c}) \to 0$ for $k \gg k_{c}.
$
In the case of Casimir effect arising due to parallel plates, the components of the wave-vector parallel to the plate surfaces remain unaffected and form a continuous spectrum; but since the field is required to vanish at the plates, the normal part is quantized in integral units of $\pi /a$: $k_{n} = n \pi /a$, where $n=0,\pm 1,\pm 2,...$. The expression for the Casimir energy per unit area, which is the change in the zero-point energy of the field, takes the following form for $d \geq 3$ dimensions \cite{greiner}:
\begin{eqnarray}
 \label{eqn:cas_en_dd1}
E_{c}^{(d)}(a, k_{c}) &=&A_d\left[ \sum_{n=-\infty}^{\infty} F(n) - \int_{-\infty}^{\infty} dn F(n) \right]\\   
 &=&2A_d\left[ \sum_{n=0}^{\infty} F(n) - \int_{0}^{\infty} dn F(n) -\frac{1}{2}F(0)\right]\nonumber   
\end{eqnarray}
where $A_d= 1/[ 2^{d+1} \pi^{(d-1)/2} \Gamma ((d-1)/2) ]$ and
\begin{equation}
F(n) =k_{c}^{d} \int_{0}^{\infty} d(y ^{2}) (y ^{2})^{\frac{(d-3)}{2}} K f (K)
\end{equation}
with
\begin{equation}
y = k_{\|}/k_{c};\quad K = \sqrt{y ^{2} + (n/ \mu)^{2}}; \quad\mu = a k_{c}/\pi.
\end{equation} 
(The $n=0$ term in the sum can be omitted since it is independent of the plate separation and does not contribute to the Casimir \textit{force} but we keep it since we do not want to throw anything away in \textit{energy density}!) In the presence of a cut-off, both terms in Eq.(\ref{eqn:cas_en_dd1}) are finite. The physically observed Casimir effect still arises from the $\textit{difference}$, since it represents the effect of the plates on the vacuum energy density. Alternatively, one might note that the presence of the plates modifies the modes that are allowed $\textit{both}$ between the plates $\textit{and}$ in the outside region. Computing the effect in a large normalization box and allowing the volume of the box to tend to infinity leads to the same effect as subtracting the continuum term. (As any condensed matter physicist will insist, the renormalization in quantum field theory is logically independent of the need for regularization.)

To illustrate the idea let us consider the case of an exponential cut-off given by $f(k/k_{c}) = \exp {(-k/k_{c})}$. With the inclusion of this cut-off, the expression for the Casimir energy per unit area can be evaluated using the Abel-Plana summation formula \cite{ictp}.
Fig.1 shows the variation of the dimensionless quantity $E_{c}^{(d)}(a, k_{c}) /k_{c}^{d}$ as a function of $\mu = a k_{c}/\pi$ (the $k_{c}$ is assumed fixed and $a$ is varied) for $d = $1 (dashed curve), 3 (thick red curve; the most relevant one for us), $d=4$ (green curve) and $d=5$ (blue curve) dimensions. The expression for $E_{c}^{(d)}(a, k_{c})$ has simple analytic forms for $d=3$ and 1. For $d=3$ we have:
\begin{equation}
 \nonumber
E_{c}^{(3)}(a, k_{c})= 
\frac{\pi^{2}\mu}{16 a^{3}}\left[ -24\mu^3 + 4 \mu^2 x + (x^{2} - 1)(2 \mu +  x) \right]
\end{equation}
where we have defined $x = \coth{(1 /2 \mu)}$, and for $d=1$:
\begin{equation}
E_{c}^{(1)} (a, k_{c}) = \frac{\pi}{2 a} \left[ \frac{e^{1/\mu}}{(e^{1/\mu}-1)^2} - \mu ^{2}\right].    \label{eqn:cas_en_d1}
\end{equation}
The $d=4,5$ cases are shown to stress that the general features are independent of $d$. 
\begin{figure}
\includegraphics[scale=1.0]{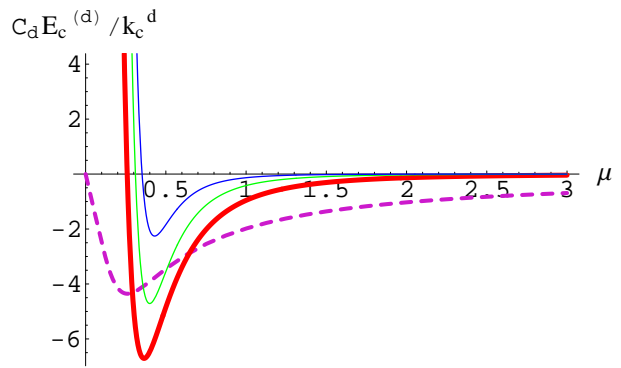}
\caption{Variation of $E_{c}^{(d)}(a, k_{c}) /k_{c}^{d}$ with $\mu = a k_{c} / \pi$ for $d=1$ (dashed curve), $3$ (thick red), $4$ (green) and $5$ (blue) spatial dimensions. Each curve is scaled by a suitable numerical factor $C_{d}$ so as to display them together; $C_{1}=50$, $C_{3}=5 \times 10^{3}$, $C_{4}=2 \times 10^{4}$ and $C_{5}=5 \times 10^{4}$.}
\label{Fig.1}
\end{figure}

Fig.1 and the analytic expressions show that for $\mu\gg1$, which corresponds to $a\gg L_c$, the curves follow the standard prediction, varying as $a^{-d}$ and the force is an attractive one. This is to be expected since the control parameter $\mu$ depends on the combination $k_{c}a$ and large $a$ corresponds to large $k_c$ and the $L_c \to 0$ limit. But the small $a$ behavior is modified, and all the curves have stable minima, which implies that at separations small compared to $L_{c}$, \textit{the force flips sign and turns repulsive!}. When $k_{c}a\to 1$ (that is, when $a\approx L_{c}$), we certainly expect to see some difference, but it might seem a bit surprising that the sign of the Casimir force changes around $a\approx L_c$. The basic reason for the turning around of the energy function is the following: the Casimir energy is essentially the difference between a sum and an integral involving the same function, as is evident from Eq.~(\ref{eqn:cas_en_dd1}). Such a difference can be expressed as an integral using the Euler-Maclaurian summation formula \cite{ictp} as
\begin{eqnarray}
\sum_{n=0}^{N} F(n) &-& \int_{0}^{N}dn F(n) -\frac{1}{2} F(0) \nonumber \\
 &=& \int_{0}^{N} \left( x - [x] - \frac{1}{2} \right) F'(x) dx 
+ \frac{1}{2}F(N) \nonumber
\end{eqnarray}
where $[x]$ denotes the greatest integer less than $x$.
From the above relation, it is clear that for monotonically increasing(decreasing) functions, the cumulative difference, as a function of $N$, always increases(decreases). [This is also easy to see geometrically by treating the sum as the sum of areas of small rectangles of unit width etc.] The function $F(n)$ we are dealing with, however, is not monotonic in the presence of a cut-off. It starts increasing with $n$, peaks around a critical value $n_{c}$ and begins to decay thereafter. For such a function, the cumulative difference initially grows more and more negative; however, beyond the critical value $n_{c}$, which corresponds to the decaying part of $F(n)$, the contribution from the sum starts dominating that of the integral, as a result of which the cumulative difference turns around. It may be noted that since in the expression for the Casimir energy, the argument of $F(n)$ is essentially the product $k L_{c}~\equiv~n L_{c} / a$, the above qualitative features would hold {\it also if} $L_{c}/a$ were varied instead, keeping $N$ fixed. From this, it follows that decreasing $a/L_{c}\propto \mu$ beyond a point would cause the variation of the Casimir energy to reverse sign, resulting in a repulsive force. This behavior is clearly independent of the dimensionality of the system.

The form of the cut-off used above was chosen because of convenience. In the absence of any sensible theory, we have no idea what $f(k/k_c)$ is, even assuming a cut-off exists. To probe the effect of a cut-off, we want to investigate a set of test functions which drop sharply but in a controllable model. These are provided by, for example, the set:

\begin{equation}
f(k/k_{c}) = \left[1+ \left(\frac{k}{k_{c}} \right) ^{\alpha}\right]^{-1}
\end{equation}
where $\alpha > (d+1)$  in order to have a finite total vacuum energy. Large values of $\alpha$ make the function drop sharply with $\alpha\to\infty$ corresponding to the Heaviside theta function.
Since the essential features do not depend on the number of dimensions, we shall, for simplicity, work in one dimension hereafter.
\begin{figure}
\includegraphics[scale=0.7]{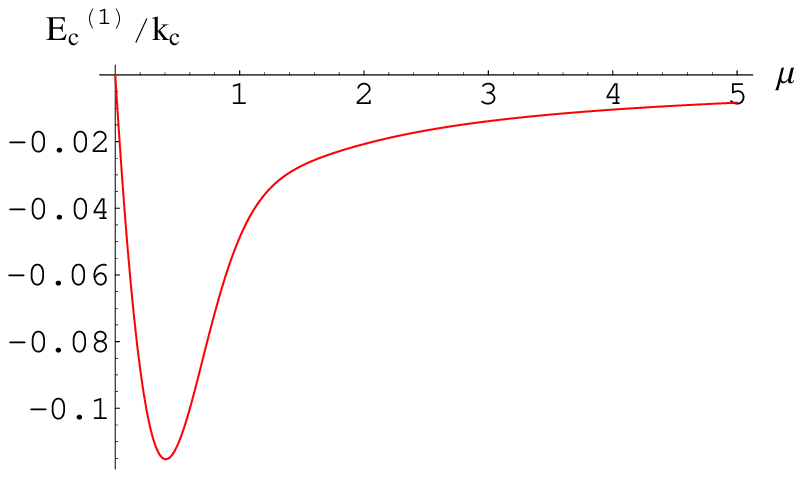}
\includegraphics[scale=0.7]{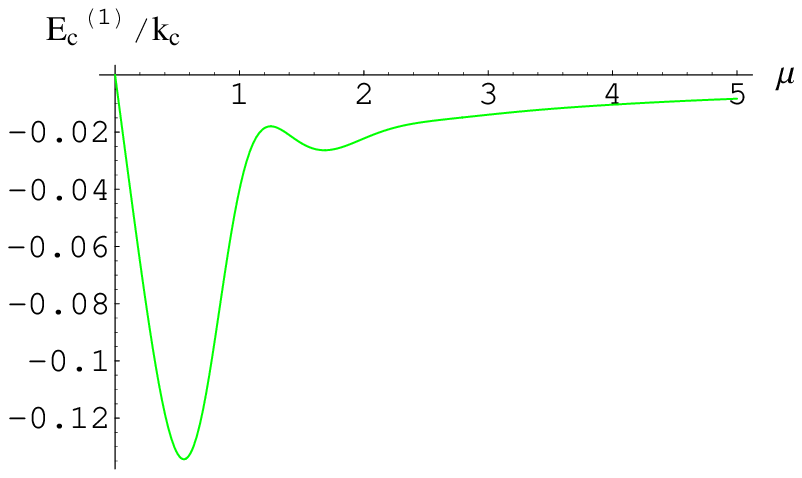}
\includegraphics[scale=0.7]{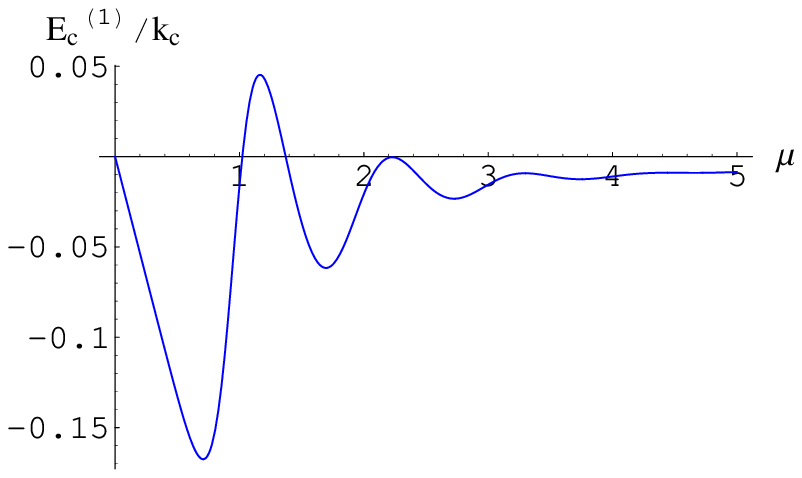}
\caption{Variation of $E_{c}^{(1)}/k_{c}$ with $\mu = a k_{c}/ \pi$ for power law cut-off functions with indices $\alpha=3, 5$ and $10$.}
\label{Fig.2}
\end{figure}
Fig.2 shows the variation of $E_{c}^{(1)} (a, k_{c})/ k_{c}$ with $\mu $ for different power law indices, evaluated numerically. The behavior for very large $a$ is unaffected by the presence of the cut-off, as expected. For large $\alpha$, the energy function undergoes several oscillations, whose number grows with the value of $\alpha$. The appearance of these oscillations is tied to the sharpness of the cut-off function but we stress that the Casimir force is well defined for all $a$ as long as the cut-off function is smooth. To see these effects more clearly, we can consider the particularly striking example of a step function cut-off (which is, of course, not smooth and corresponds to $\alpha\to\infty$) given by $f(k/k_{c}) = \Theta (k - k_{c})$. In this case the energy expression is
\begin{equation}
E_{c}^{(1)} (a, k_{c}) = \frac{\pi}{2 a} \left( \left[\mu \right]^{2}  + \left[\mu \right] - \mu ^{2} \right)
\end{equation}
where $[x]$ denotes the greatest integer less than $x$. The plot for this case is shown in Fig.3. Now the force is discontinous, since the cut-off is not smooth. It is obvious that the oscillations are related to the sharpness of the cut-off function.

In all these cases as well, 
the Casimir force turns repulsive for small separations.
In fact, using the Abel-Plana summation formula one can obtain an asymptotic expression for small separations (corresponding to $\mu \ll 1$):
\begin{equation}
E_{c}^{(1)} (a, k_{c})  \stackrel{\mu \to 0}{\longrightarrow} - \frac{a k_{c}^{2}}{\alpha} \csc \left( \frac{2 \pi}{\alpha} \right); \quad\alpha>2
\end{equation}
which clearly leads to a repulsive force.
The above results suggest that, in the presence of a cut-off length $L_{c}$ for zero-point vacuum fluctuations, the Casimir force between parallel plates turns repulsive for  $a\lesssim L_{c}$, and this feature is independent of the functional form of the cut-off.
The above analysis would remain valid even for the case of an electromagnetic field in 3+1 dimensions (apart from an extra factor of 2 in all the relevant expressions arising due to the fact that there are 2 polarization states for photons) \cite{greiner}.
\begin{figure}
\includegraphics[scale=0.7]{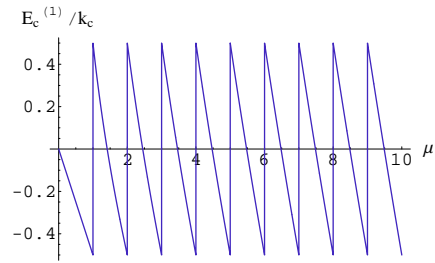}
\caption{Variation of $E_{c}^{(1)}/k_{c}$ with $\mu = a k_{c} / \pi$ for the step function cut-off.}
\label{Fig.3}
\end{figure}

If the zero-point energy causing the Casimir effect is indeed responsible for the non-zero value of the cosmological constant \cite{cc}, there must exist a natural UV cut-off that renders the total vacuum energy finite, leading to the observed value of $\Lambda$. To agree with the currently observed value for the dark energy density \cite{cc}, this cut-off scale must be of the order of 0.1 mm. Our results clearly  indicate that, if such a cut-off does actually exist in nature, the Casimir force would be repulsive for plate separations smaller than the cut-off scale ($\sim 0.1 mm$). Experimentally, the Casimir force between metallic surfaces due to the electromagnetic field has been measured up to length scales of  $\sim 100nm$, and the results seem to agree well with the standard predictions \cite{expts}. This implies that any possible cut-off must be less than 100 nm, which corresponds to an energy density of at least $\sim 1 (eV)^{4}$. The huge difference of this value compared with $\rho_{DE}$ makes it improbable that the zero-point energy of the vacuum, made finite with a cut-off, contributes to the cosmological constant. The striking discrepancy of the expected deviations of the Casimir force in the presence of a millimeter cut-off with the experimental results, as revealed by our analysis, makes it unclear if there is really any need to look for such a cut-off through other experiments \cite{beck}.

We can summarise the situation as folllows. If the dispersion relation for zero-point energy modes have a form $\omega=kf(k/k_c)$ then the zero-point energy will be finite and will be about $\rho_V\approx k_c^{4}$. At the same time, it will modify the form of the Casmir force between parallel plates drastically for $ak_c\lesssim 1$. (These are, of course, different but related effects; $\rho_V$ arises from the integral over the modes while the Casimir effect arises from the difference between the zero-point energies in the presence and absence of plates.) From the laboratory data (for $a_L\approx 100 nm$), we can conclude that $k_c\gtrsim a_L^{-1}$. If these modes couple to gravity, the resulting vacuum energy density will be at least about $10^{12}$ times larger than the observed value. Hence it is mandatory that somehow gravity does not couple to these bulk vacuum modes, which is --- of course --- the well known cosmological constant problem. It is, however, possible to achieve this in models of gravity which are holographic and the degrees of freedom live on the surface of the region. In such models, the transformation $T_{ab}\to T_{ab}+\Lambda g_{ab}$ is an invariance of the theory and gravity does not couple to bulk cosmological constants directly. The cosmological constant arises as an integration constant to the equations of motion \cite{tpholo} in a manner similar to the way in which the mass $M$ arises in the solution to spherically symmetric vacuum Einstein's equations.

Finally, we note that the minimum for $E_{c}$ as a function of the separation might have some implications for reducing the size of compact
extra dimensions. If the cut-off is at Planck length, it may be possible to arrange matters such that the regularized energy density has a minimum value as a function of the radii of the compact dimensions at about Planck length. This possibility is under investigation.

This study was initiated from a question raised by our colleague K. Subramanian, whom we thank for extensive discussions and comments on the draft of the paper. G.M. and S.S. are supported by the Council of Scientific \& Industrial Research, India.


\begin{thebibliography}{20}


\bibitem{cas}
 H. B. G. Casimir, {\it Proc. Kon. Ned. Akad.} {\bf 51}, 793 (1948).

\bibitem {milonni}
P. W. Milonni, {\it The Quantum Vacuum - An Introduction to Quantum Electrodynamics} (Academic, New York, 1994).

\bibitem {milton}
K. A. Milton, arXiv:hep-th/0009173; K. A. Milton, arXiv:hep-th/0406024.

\bibitem {greiner}
G. Plunien, B. Muller and W. Greiner, Phys. Rep. {\bf 134}, 87 (1986).

\bibitem{expts}
U. Mohideen and A. Roy, Phys. Rev. Lett. {\bf 81}, 4549 (1998); S. K. Lamoreaux, Phys. Rev. Lett. {\bf 78}, 5 (1997); G. Bressi et al., Phys. Rev. Lett. {\bf 88}, 4 (2002).

\bibitem{minlenpap}  See e.g.
T. Padmanabhan, Phys. Rev. Lett. {\bf 78},  1854 (1997) (arXiv: hep-th/9608182);
Phys. Rev. Lett. \textbf{60}, 2229  (1988);
S. Corley, T. Jacobson, Phys. Rev. {\bf D} \textbf{54}, 1568 (1996);
J. C. Niemeyer, R. Parentani, Phys. Rev. {\bf D} \textbf{64} 101301 (2001);
 J. Kowalski-Glikman,  Phys. Lett. B \textbf{499} 1 (2001);
 G. Amelino-Camelia, Int. J. Mod. Phys. D \textbf{11} 35 (2002).
For a review, see L. J. Garay, Int. J. Mod. Phys. A \textbf{10}, 145 (1995).

\bibitem{nouc}
K. Nouicer, J. Phys. A {\bf 38}, 10027-10035 (2005) (arXiv:hep-th/0512027); U. Harbach and S. Hossenfelder, Phys. Lett. B {\bf 632}, 379-383 (2006).

\bibitem{beck}
C. Beck, arXiv:astro-ph/0512327.

\bibitem{cc}
C. L. Bennett et al., Astrophys. J. Supp. Series {\bf 148}, 1 (2003) (arXiv:astro-ph/0302207);
 P. Astier et al., arXiv:astro-ph/0510447;
H.K. Jassal et al., \textit{MNRAS} \textbf{356}, L11-L16 (2005), (arXiv:astro-ph/0404378);
             (arXiv:astro-ph/0601389).
For earlier evidence, see, 
G.~Efstathiou et al.,  Nature,  (1990), \textbf{348}, 705;
J.~S.~Bagla, T.~Padmanabhan and J.~V.~Narlikar, Comments   on Astrophysics,  (1996), \textbf{18}, 275 [astro-ph/9511102].
For a review, see 
T. Padmanabhan, Phys. Rep. {\bf 380}, 235 (2003).

\bibitem{ictp}
J. P. Dowling, {\it The Mathematics of the Casimir Effect}, IC/87/41 (1987).

\bibitem{table}
I. S. Gradshteyn and I. M. Ryzhik, {\it Table of Integrals, Series \& Products} (Academic Press, New York, 1989).

\bibitem{tpholo}
T. Padmanabhan, Int. J. Mod. Phys. D {\bf 14}, 2263-2270 (2005) (arXiv:gr-qc/0510015); T. Padmanabhan, \textit{Dark Energy: Mystery of the Millennium }  (arXiv:astro-ph/0603114); arXiv:gr-qc/0412068.

\end{thebibliography}
\end{document}